\documentclass[%
 aip,
 amsmath,amssymb,
 reprint,%
]{revtex4-1}

\usepackage{graphicx}
\usepackage{dcolumn}
\usepackage{bm}
\usepackage{color}

\usepackage[ngerman,english]{babel} 
\usepackage[utf8]{inputenc}
\usepackage[T1]{fontenc}
\usepackage{mathptmx}
\usepackage{upgreek}

\usepackage{floatrow}    
\newfloatcommand{capbtabbox}{table}[][\FBwidth]

\usepackage{subfig}
\usepackage{caption}

\usepackage{float}

\begin{document}

\preprint{AIP/123-QED}

\title{Flicker and random telegraph noise between gyrotropic and dynamic C-state of a vortex based spin torque nano oscillator}

\author{Steffen Wittrock}\email[]{steffen.wittrock@cnrs-thales.fr}\affiliation{Unit\'{e} Mixte de Physique CNRS, Thales, Univ. Paris-Saclay, 1 Avenue Augustin Fresnel, 91767 Palaiseau, France }
\author{Philippe Talatchian}\altaffiliation[Present address: ]{Institute for Research in Electronics and Applied Physics, Univ. of Maryland, College Park,  20899-6202, MD, USA }\affiliation{Unit\'{e} Mixte de Physique CNRS, Thales, Univ. Paris-Saclay, 1 Avenue Augustin Fresnel, 91767 Palaiseau, France }
\author{Miguel Romera}\altaffiliation[Present address: ]{GFMC, Departamento de Física de Materiales, Universidad Complutense de Madrid, 28040 Madrid, Spain}\affiliation{Unit\'{e} Mixte de Physique CNRS, Thales, Univ. Paris-Saclay, 1 Avenue Augustin Fresnel, 91767 Palaiseau, France }
\author{Mafalda Jotta Garcia}\affiliation{Unit\'{e} Mixte de Physique CNRS, Thales, Univ. Paris-Saclay, 1 Avenue Augustin Fresnel, 91767 Palaiseau, France }
\author{Marie-Claire Cyrille}\affiliation{ Univ. Grenoble Alpes, CEA-LETI, MINATEC-Campus, 38000 Grenoble, France }
\author{Ricardo Ferreira}\affiliation{ International Iberian Nanotechnology Laboratory (INL), 471531 Braga, Portugal }
\author{Romain Lebrun}\affiliation{Unit\'{e} Mixte de Physique CNRS, Thales, Univ. Paris-Saclay, 1 Avenue Augustin Fresnel, 91767 Palaiseau, France }
 \author{Paolo Bortolotti}\affiliation{Unit\'{e} Mixte de Physique CNRS, Thales, Univ. Paris-Saclay, 1 Avenue Augustin Fresnel, 91767 Palaiseau, France }
 \author{Ursula Ebels}\affiliation{ Univ. Grenoble Alpes, CEA, INAC-SPINTEC, CNRS, SPINTEC, 38000 Grenoble, France }


\author{Julie Grollier}\affiliation{Unit\'{e} Mixte de Physique CNRS, Thales, Univ. Paris-Saclay, 1 Avenue Augustin Fresnel, 91767 Palaiseau, France }
 \author{Vincent Cros}\email[]{vincent.cros@cnrs-thales.fr}\affiliation{Unit\'{e} Mixte de Physique CNRS, Thales, Univ. Paris-Saclay, 1 Avenue Augustin Fresnel, 91767 Palaiseau, France }

\date{\today}

\begin{abstract}
Vortex based spin torque nano oscillators (STVOs) can present more complex dynamics than the spin torque induced gyrotropic (G) motion of the vortex core.  
The respective dynamic modes and the transition between them can be controlled by experimental parameters such as the applied dc current. 
An interesting behavior is the  stochastic transition from the G- to  a dynamic C-state occurring for large current densities. Moreover, the C-state oscillations exhibit a constant active magnetic volume. 
We present noise measurements in the different dynamic states that allow accessing  specific properties of the stochastic transition, such as the characteristic state transition frequency. 
Furthermore, we confirm, as theoretically predicted, an increase of flicker noise with $I_{dc}^2$ when the oscillation volume remains constant with the current.  
These results bring insight into the potential optimization of noise properties sought for many potential rf applications with spin torque oscillators. 
Furthermore, the investigated stochastic characteristics open up new potentialities, for instance in the emerging field of neuromorphic computing schemes. 
\end{abstract}

\maketitle


\section{Introduction}

Spin torque nano oscillators (STOs) exploit magnetoresistive effects in order to convert magnetization dynamics into rf electrical signals. 
Exhibiting interesting properties, such as a nanometric size, complete CMOS compatibility\cite{Makarov2016}, radiation hardness\cite{Hughes2012} and high frequency tunability\cite{Arun2020}, they are considered the next-generation devices for rf applications and communications\cite{Locatelli2013,Ebels2017}. 
However, their potentialities go even beyond this and include exploitation for hard disk reading\cite{Sato2012}, or, based on the STOs' high responsiveness to external stimuli, broadband frequency  detection\cite{Litvinenko2020} and microwave energy harvesting\cite{Fang2019}. 
Furthermore, along with the possibility to efficiently synchronize two or more STOs together\cite{Kaka2005,Lebrun2017,Tsunegi2018} in order to leverage novel neuro-inspired computing architectures for e.g. pattern recognition\cite{Romera2018}, also more complex phenomena have been demonstrated in STOs, such as stochastic \cite{Jenkins2019,Wittrock2021_C-state-APL} or even chaotic\cite{Petit-Watelot2012,Devolder2019} behavior. 

It has recently been shown\cite{Wittrock2021_C-state-APL} that the transition between two precession modes in a vortex based STO (STVO) exhibits stochastic characteristics that can be controlled by the external parameter, i.e. the dc current or applied magnetic field. 
The two involved modes are the vortex core gyrotropic state and beyond, the dynamic C-state motion of the magnetization distribution with particularly  interesting characteristics, such as a nearly constant magnetic oscillation volume\cite{Wittrock2021_C-state-APL}. 

So far, the main limitation of all the above mentioned possible STO applications is their large noise characteristics, which is necessary to be reduced and better understood. 
Flicker noise, that is dominant at longer timescales of sought stability, has been studied and classified mainly in the gyrotropic regime\cite{Wittrock2019_PRB,Wittrock2020-SciRep} of a STVO. 
We recently found in particular  a dependence of the noise level on the active magnetic volume $V$ of the oscillation, along with the nonlinear dynamical parameters\cite{Wittrock2019_PRB}, and predict a Hooge-like increase\cite{Hooge1969} with $I_{dc}^2$ when the volume $V$ is constant. 
In this general sense, the described C-state oscillations with constant oscillation volume $V$ represent an interesting case to study the system's flicker noise level. 
Furthermore, along with the characterization of the flicker noise in the C-state, we here perform noise measurements in the transition between gyrotropic and C-state. 
Typical for a two-level stochastic system, we demonstrate the occurrence of  random telegraph noise (RTN) within the transition, which turns out to be useful in order to better characterize the stochastic transition.


\section{Experiment}

\begin{figure*}[bth!]
  \centering  
  \captionsetup[subfigure]{labelformat=empty}
\subfloat[ \label{fig_C-state:spectra_and_sketch_520mT}] 
  { 
  \includegraphics[width=0.3\textwidth]{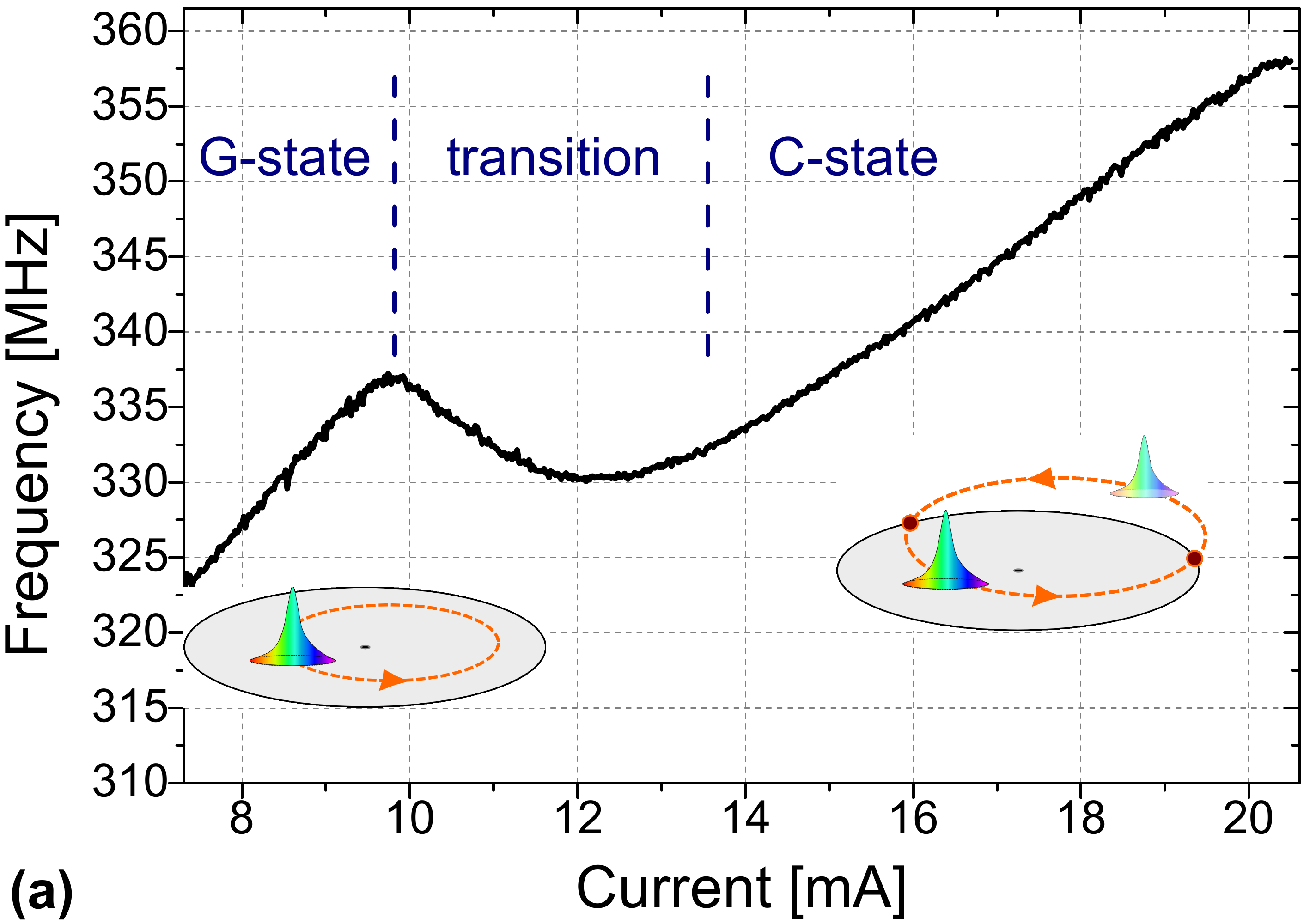}  } 
 \subfloat[ \label{fig_C-state:PN_17mA}] 
  { 
  \includegraphics[width=0.305\textwidth]{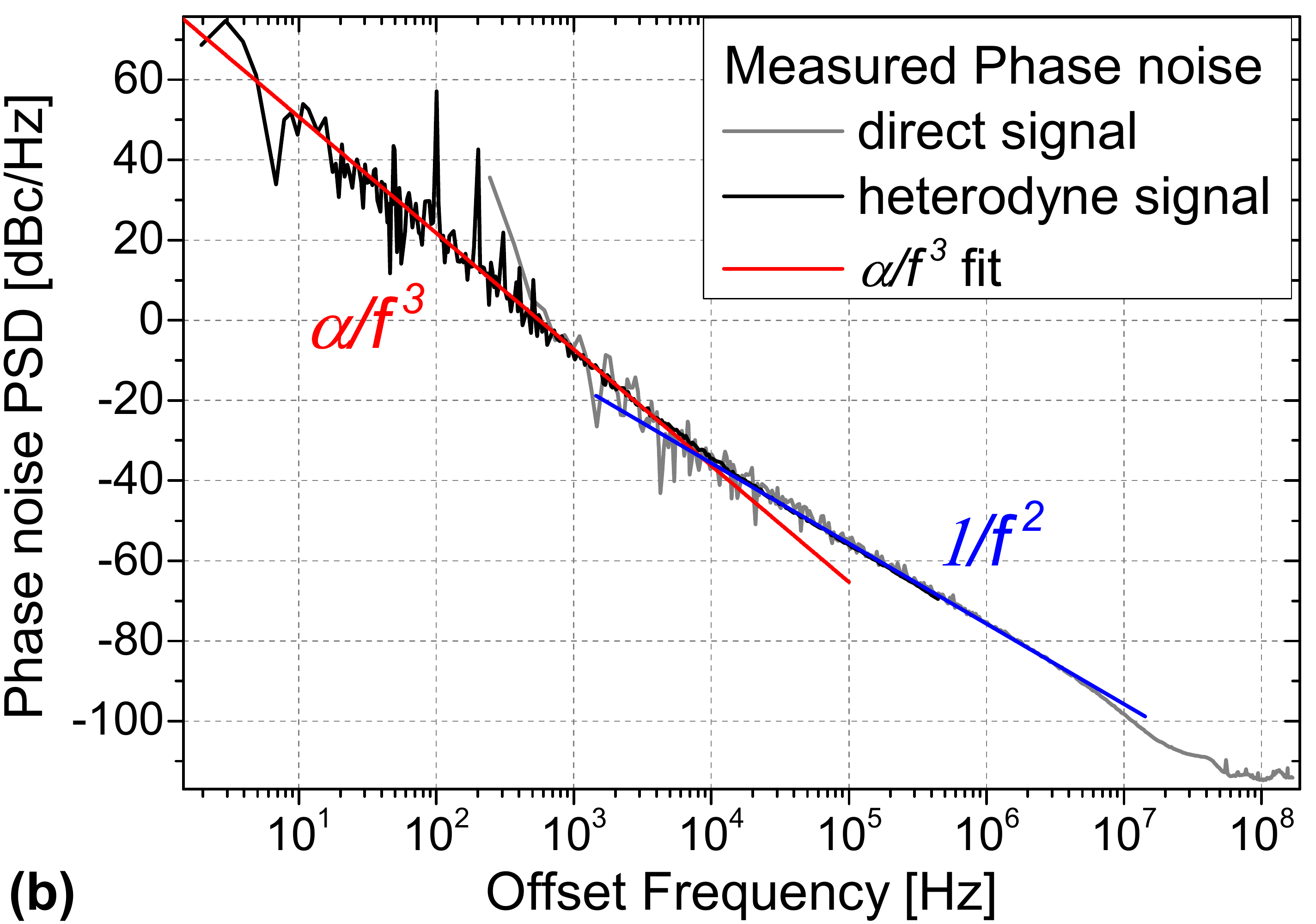}  } 
 \subfloat[ \label{fig_C-state:flicker-prefactor}] 
  { 
  \includegraphics[width=0.31\textwidth]{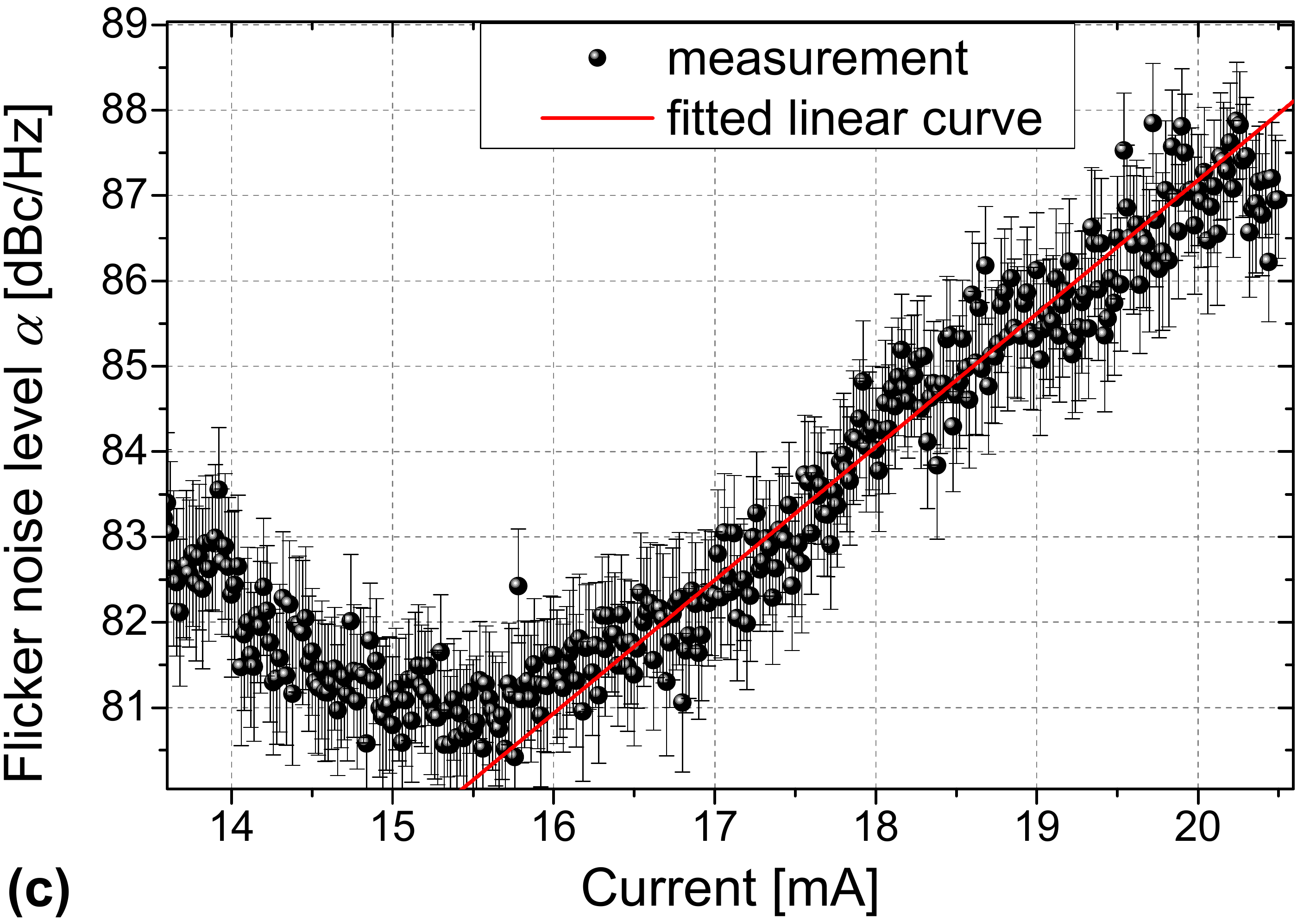}  } 
  \caption[]{ (a) Measured frequency vs. current evolution of the described G- and C-state regime with sketched motion of the vortex core inside/outside the nanodisk characterizing G- and C-state. (b) Phase noise curve in the C-state at $17\,$mA with fitted $\alpha/f^3$ flicker noise contribution. (c) Corresponding flicker noise level $\alpha$ vs. dc current in the C-state regime. The applied perpendicular-to-plane magnetic field is $\mu_0 H_{\perp} = 520\,$mT.  }
  \label{fig_C-state:flicker-noise}
\end{figure*}

The studied samples are the same as those used in Ref. \cite{Wittrock2021_C-state-APL}, i.e. circularly shaped nanopillars of $300\,$nm diameter consisting of a  synthetic antiferromagnetic stack (SAF) as a pinning layer. 
In order to exploit the tunnel magnetoresistive (TMR) effect, a MgO tunnel barrier separates the SAF and a $7\,$nm NiFe free layer in the magnetic vortex configuration.
The TMR ratio lies around $50\,$\% at room temperature and the complete layer stack is \selectlanguage{ngerman} SAF/""MgO($1$)/""Co$_{40}$Fe$_{40}$B$_{20}$($1.5$)/""Ta($0.2$)/""Ni$_{80}$Fe$_{20}$(7)/""Ta(5)/Ru(7) \selectlanguage{english} with the SAF composed of \selectlanguage{ngerman} PtMn($20$)/""Co$_{70}$Fe$_{30}$($2$)/""Ru($0.85$)/""Co$_{40}$Fe$_{40}$B$_{20}$($2.2$)/""Co$_{70}$Fe$_{30}$($0.5$) \selectlanguage{english} and the layer thicknesses in brackets.  
The TMR effect converts the spin torque driven vortex dynamics into a rf electrical signal which is simultaneously measured on a spectrum analyzer and an oscilloscope. 
A  perpendicular-to-plane magnetic field $\mu_0 H_{\perp}$ tilts the magnetization of the SAF's top layer in order to create a perpendicular-to-plane spin polarization providing the necessary spin torque component to reach self-sustained dynamics\cite{Dussaux2012}.  

Noise data are gathered from single-shot oscilloscope voltage time traces and processing via the Hilbert transform method\cite{Bianchini2010,Quinsat2010,Wittrock2019_PRB}. 
To obtain noise data at low offset frequencies close to the carrier and furthermore resolve small frequency variations, 
 a heterodyne detection technique\cite{Keller2010,Eklund2014,Wittrock2019_PRB} (signal down-conversion via high-side injection and low-pass filtering (bandwidth DC to $22\,$MHz)) is used. 
 

\section{Flicker noise in the C-state}


The dynamic C-state describes the motion of the magnetic vortex core in a circular disk beyond the gyrotropic (G) state. 
At one point in the periodic trajectory, the vortex core reaches the disk boundary, disappears and an in-plane magnetic C-state distribution forms and precesses for a part of the oscillation period. 
We have recently shown\cite{Wittrock2021_C-state-APL} that this behavior is mainly triggered by the presence of a field-like torque and can be identified by a rather abrupt decrease of the oscillation frequency at larger applied dc currents $I_{dc}$. 
In fig. \ref{fig_C-state:flicker-noise}, a characteristic measurement with a perpendicular-to-plane magnetic field $\mu_0 H_{\perp} = 520\,$mT is shown: Recognized from the frequency vs. current curve (fig. \ref{fig_C-state:spectra_and_sketch_520mT}),  the STVO at smaller dc currents oscillates in the G-state up to a transition from  G- to C-state which stabilizes above $\sim 14\,$mA. 
In fig.  \ref{fig_C-state:spectra_and_sketch_520mT}'s inset, a sketch of the vortex core trajectories is presented.

In fig. \ref{fig_C-state:PN_17mA}, we show a phase noise measurement\cite{Bianchini2010,Quinsat2010,Wittrock2019_PRB} in the C-state, recorded at room temperature at $I_{dc}=17\,$mA. 
At higher frequency offsets, a $1/f^2$ noise characteristics due to thermal fluctuations is observed \cite{Tiberkevich2008} while at lower frequency offsets a $1/f^3$ flicker noise process is detected\cite{Wittrock2019_PRB}. 
Performing $\alpha /f^3$ flicker noise fits, we present in fig. \ref{fig_C-state:flicker-prefactor} the evolution of the flicker noise level $\alpha$ as a function of the applied dc current $I_{dc}$. 
A flicker noise minimum is reached at $I_{dc} \approx 15.5\,$mA, where we assume the C-state to be stabilized after the transition. 
For larger applied currents, the noise level indeed increases. 
Performing a linear fit on the curve as depicted in fig. \ref{fig_C-state:flicker-prefactor}, we find a curve slope of $\sim (1.6 \pm 0.1)\,$dBc/(Hz$\cdot$mA). 
The Hooge formula\cite{Hooge1969,Fermon2013} $\alpha \sim \alpha_H I_{dc}^2/V$ predicts a value of 2  (note the logarithmic scale) at constant active volume\cite{Wittrock2019_PRB}. 
 The fact that we do not find an increase with slope of $2$ but slightly lower might be explained by the complex current dependence of the Hooge-parameter $\alpha_H$, a sort of quality factor of the system\cite{Hooge1969},  which usually decreases for higher applied bias\cite{Gokce2006,Almeida2008,Aliev2007,Wittrock2019_PRB} linked to the simultaneously decreasing TMR value. 
 Furthermore, the dynamical parameters of the oscillation, such as the damping rate back to the limit cycle, might play a role\cite{Wittrock2019_PRB,Slavin2009}. 
In general, in the gyrotropic state the flicker noise level is decreasing with increasing current $I_{dc}$ \cite{Wittrock2019_PRB}. 
Our measurement confirms that once the oscillation volume saturates, what is the case also in the G-state at high currents, the flicker noise level evolution however turns and increases with increasing $I_{dc}$, hence the Hooge formula description dominates the evolution.   
Thus, it means that there is a minimum in the noise level which can be exploited in order to minimize the noise level in the STO operation.

\section{Random Telegraph noise in the state transition}

\begin{figure*}[bth!]
  \centering  
  \captionsetup[subfigure]{labelformat=empty}
\subfloat[ \label{fig_C-state:spectra_RTN}] 
  { 
  \includegraphics[width=0.288\textwidth]{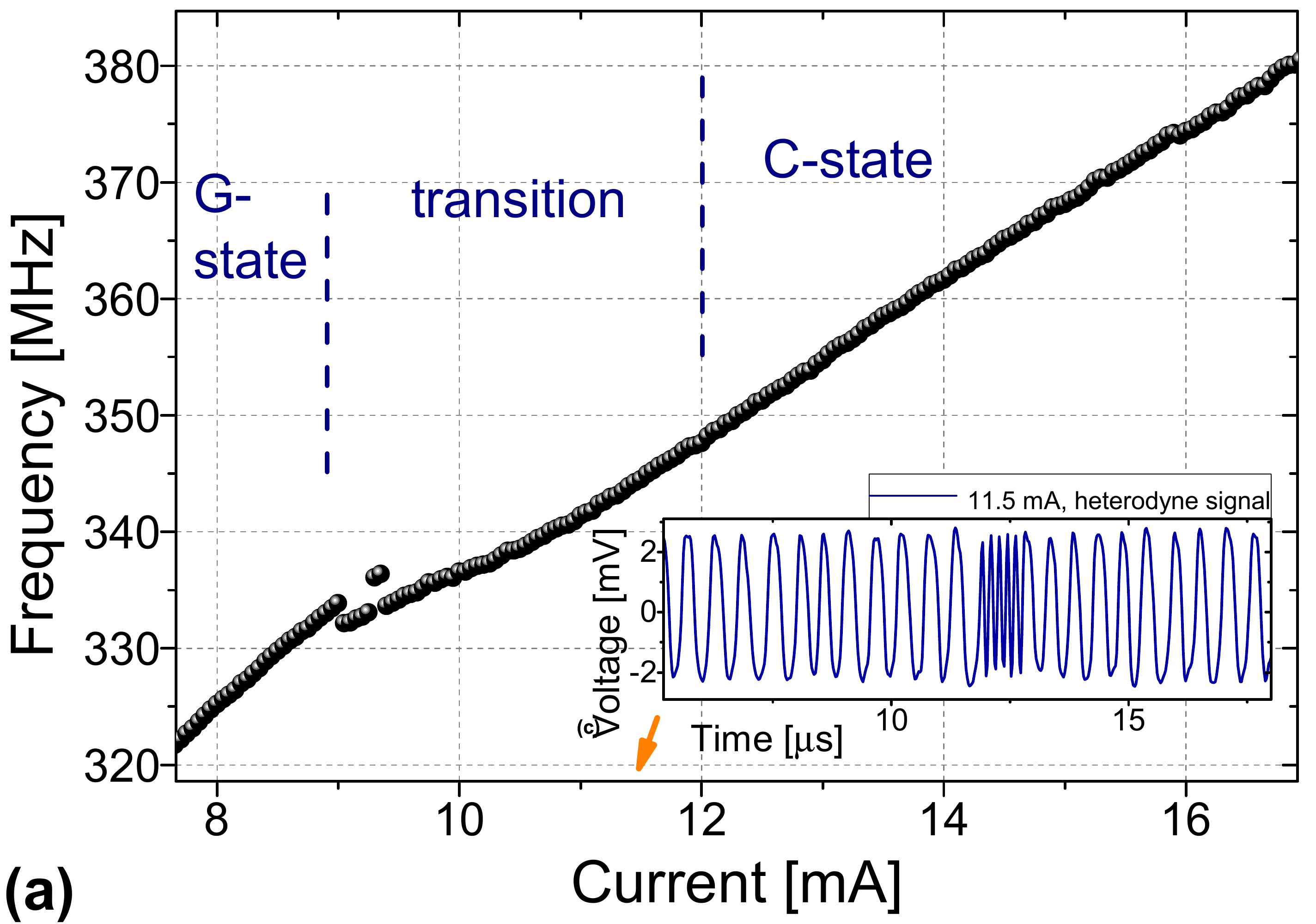}  } 
 \subfloat[ \label{fig_C-state:noisePSD_RTN}] 
  { 
  \includegraphics[width=0.287\textwidth]{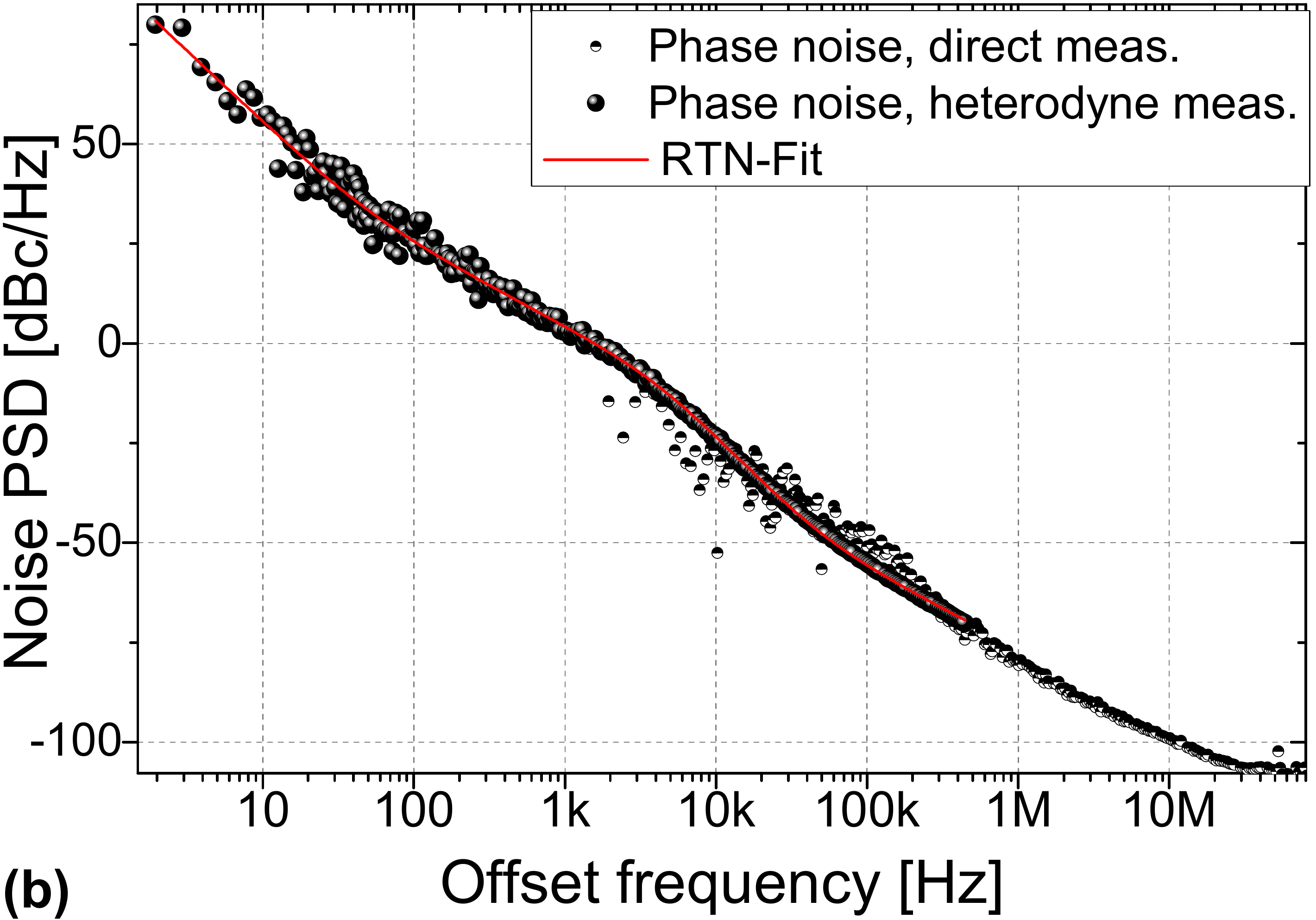}  } 
   \subfloat[ \label{fig:RTN-fit-parameters}] 
  { 
  \includegraphics[width=0.294\textwidth]{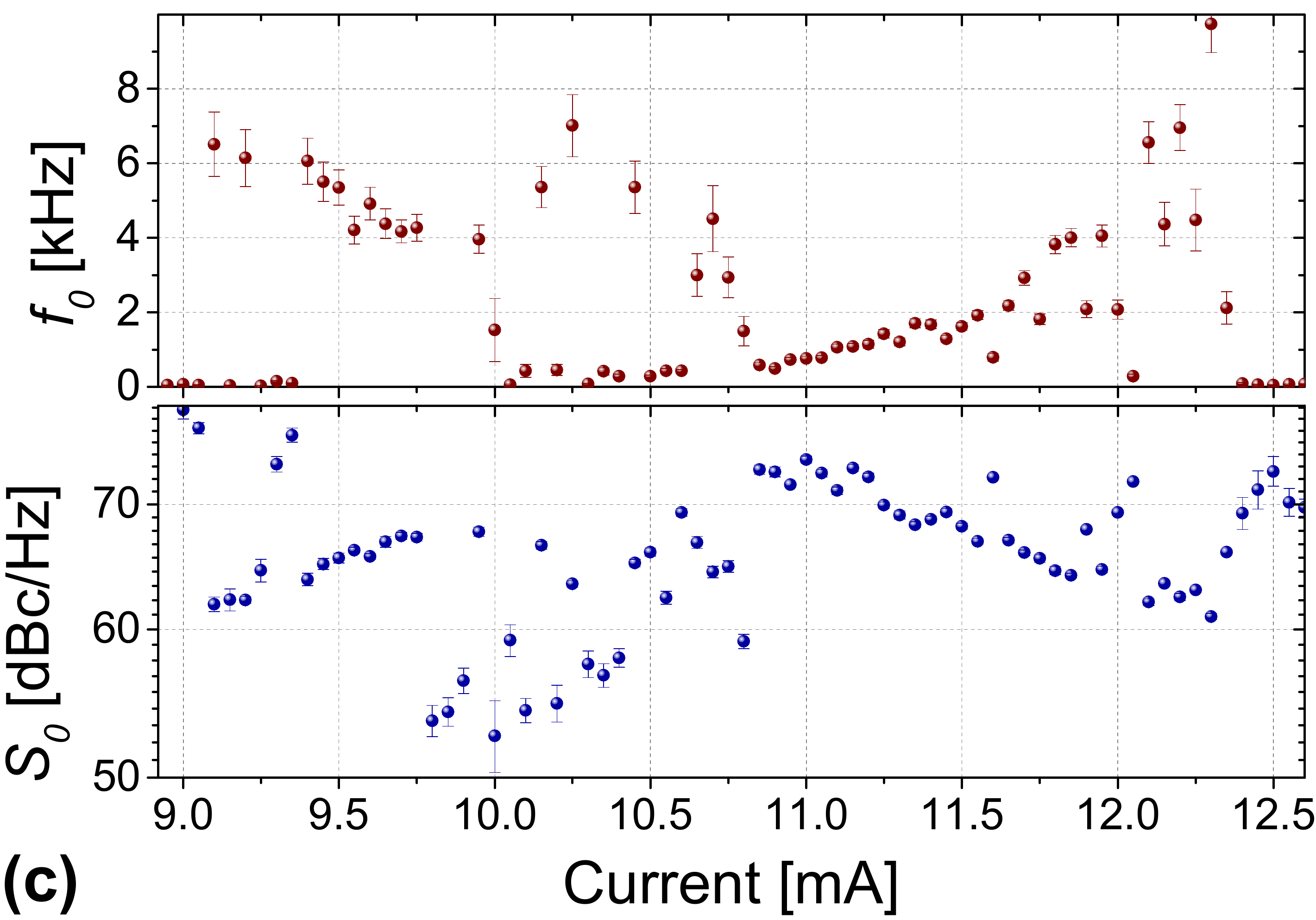}  } 
  \caption[]{ (a) Via spectrum analyzer measured frequency  vs. $I_{dc}$ with stochastic G$\rightarrow$C-state transition.  The inset shows a corresponding time trace of the heterodyne signal at $I_{dc}=11.5\,$mA highlighting the temporal change of state frequency, here in the interval $\sim [11.8;12.8]\,\upmu$s. The frequency difference of this interval to the rest of the shown curve is about $8\,$MHz.  
  (b) Noise PSD with RTN fit on the phase noise data at $I_{dc}=11.8\,$mA. 
  (c) Parameters of the RTN fits as a function of the current. 
   The applied perpendicular-to-plane magnetic field is $\mu_0 H_{\perp} = 425\,$mT. 
   }
  \label{fig_C-state:RTN}
\end{figure*}



In fig. \ref{fig_C-state:spectra_RTN}, we show the frequency vs. current evolution of a (different) measurement at $\mu_0 H_{\perp} = 425\,$mT showing a transition from G- to C-state from $\sim 9\,$mA on. 
With similar characteristics of G- and C-state as before, the frequency decrease in comparison seems rather abrupt. 
However,  traces from both of the two states are detected in the entire range from $8.9\,$mA up to $12.5\,$mA. 
At $\sim 9.3\,$mA, inside the transition regime, the G-state oscillation is even stable  at least for the duration of the measurement ($\sim 10\,$ms). 
However, within the transition the system is constantly changing between the two states with frequency differences of typically a few MHz. 
This is  represented in the heterodyne voltage signal in fig. \ref{fig_C-state:spectra_RTN}'s inset with the local oscillator frequency chosen such that it is $\sim 2\,$MHz larger than the determined principal frequency shown in fig. \ref{fig_C-state:spectra_RTN}. 
At $\sim 11.8\,\upmu$s, the oscillator changes from $343\,$MHz  into another state of $8\,$MHz higher frequency for a couple of periods. 
 By performing phase noise measurements inside the transition (fig. \ref{fig_C-state:noisePSD_RTN}), we observe a noise PSD which is no more solely described by only thermal and flicker contributions. 
 Instead, an additional random telegraph noise (RTN) process with a typical characteristic offset frequency $f_0$ of a few kHz adds to the noise PSD in fig. \ref{fig_C-state:noisePSD_RTN}. 
 The appearance of such RTN contribution to the noise is found in a large range of $I_{dc}$  from $\sim 8.9\,$mA to $\sim 12.5\,$mA and corresponds to the bistability of the two states, i.e. a change of frequency as also presented in fig. \ref{fig_C-state:spectra_RTN}'s inset. 
 This type of noise arises from fluctuations between two states\cite{Fermon2013}. 
In order to obtain characteristic parameters of the two level time signal, we perform fits to the noise data according to\cite{Almeida2008}
\begin{align*}
S_{\delta \phi} = 10\cdot \log \left( \frac{\alpha}{f^{\beta}} + \frac{S_0}{\left(1 + (f/f_0)^2\right) f^2}     \right) ~~~,
\end{align*}
\hspace{0.25cm} as represented in fig. \ref{fig_C-state:noisePSD_RTN}. 
Here, $\alpha/f^{\beta}$ with $\beta\approx 3$ reflects the flicker noise at low offset frequencies. 
The second term describes the RTN contribution with $S_0$ the RTN amplitude and $f_0$ the characteristic RTN frequency, both chosen as fitting parameters.


\begin{figure}[bth!]
  \centering  
   \captionsetup[subfigure]{labelformat=empty}
\subfloat[ \label{fig_C-state:f_vs_t}] 
  { 
  \includegraphics[width=0.72\textwidth]{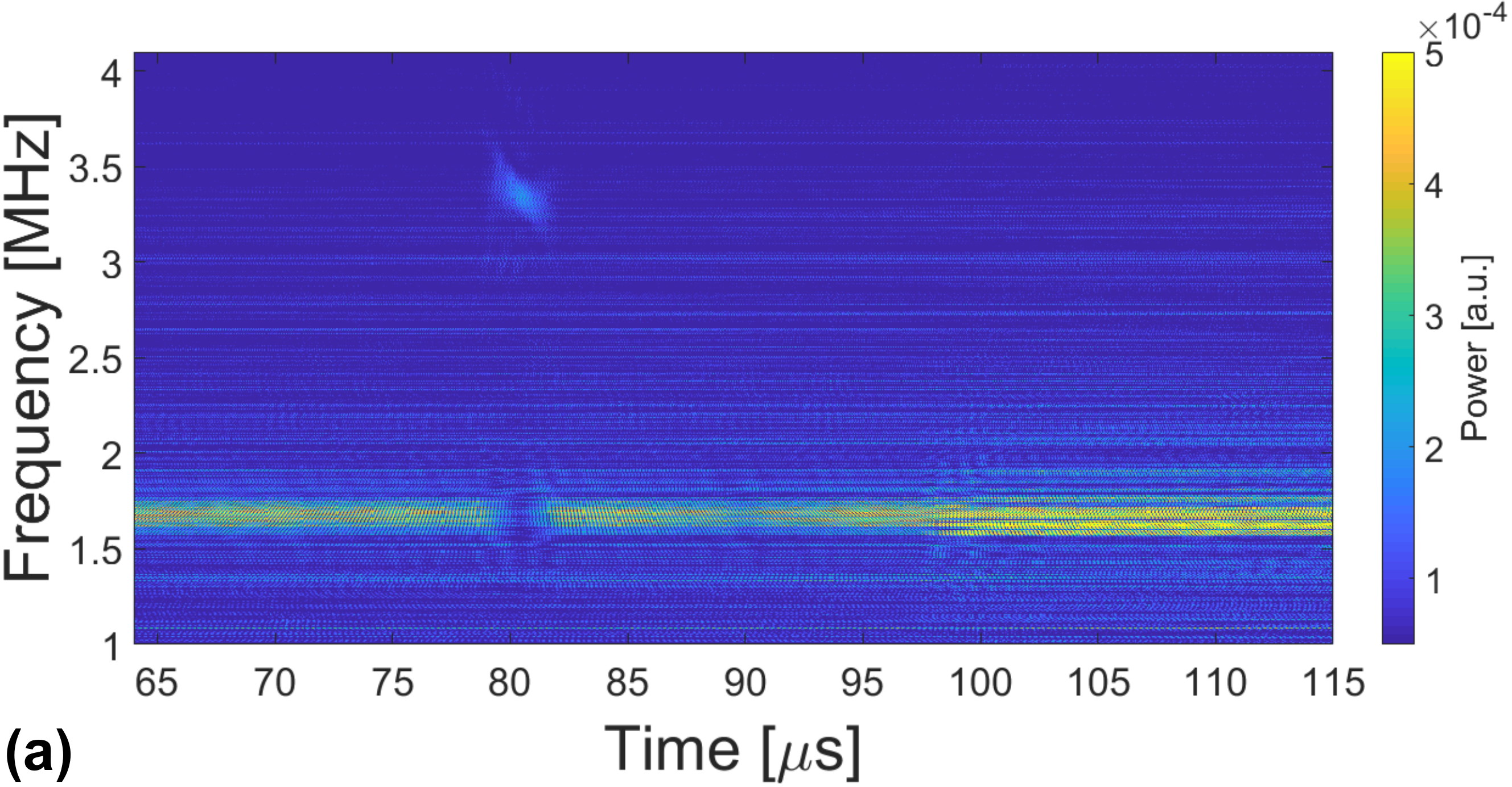}  } 
  
  \centering
 \subfloat[ \label{fig_C-state:probability}] 
  { 
  \includegraphics[width=0.58\textwidth]{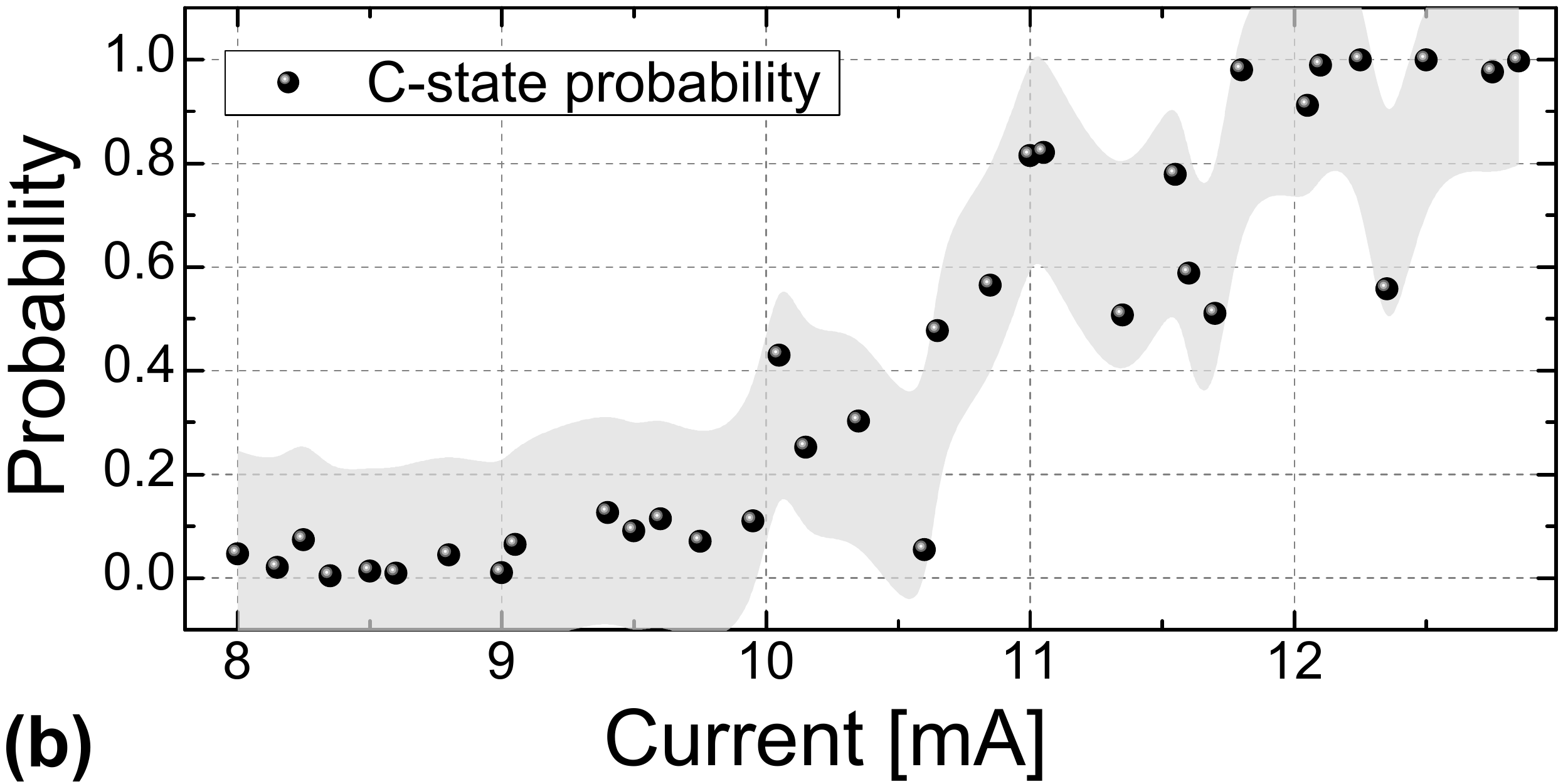}  } 
  \caption[]{ (a) Wigner-Ville time-frequency spectrogram at $I_{dc}= 10.25\,$mA. 
  (b)   Determined C-state probability vs. current through the G- to C-state transition regime from $\sim 8.9\,$mA to $\sim 12.5\,$mA. 
   Data correspond to those shown in fig. \ref{fig_C-state:RTN}. 
   }
  \label{fig_C-state:f_vs_t_probability}
\end{figure}

The obtained RTN parameters as a function of the applied current are shown in fig. \ref{fig:RTN-fit-parameters}. 
The characteristic frequency $f_0$ reveals that the stochastic state switching events occur on a kHz scale. 
It firstly decreases with increasing $I_{dc}$ for $I_{dc}\lesssim 10.5\,$mA and subsequently increases in the range $[10.6;12.5]\,$mA  until the C-state oscillation becomes stable for $I\gtrsim 12.5\,$mA. 
Note that in the range $[10;10.6]\,$mA, the characteristic frequency is partly too low for a proper parameter determination and furthermore, the frequency of the local oscillator might be between the states' frequency levels making them difficult to distinguish. 
Along with the probability to be in the G- or C-state, also the switching speed represented by $f_0$ can therefore be manipulated  by the applied current, additionally adding a further system handling. 
In fig. \ref{fig_C-state:f_vs_t_probability}, we analyze the state probability inside the transition more in detail. 
By performing Wigner-ville time-frequency spectra\cite{Cohen1995-book} of the heterodyne signal, such as shown in fig. \ref{fig_C-state:f_vs_t} for $I_{dc}=10.25\,$mA, the frequency and hence the state probability can be analyzed. 
In fig. \ref{fig_C-state:f_vs_t}, it is clearly seen that at $\sim 80\,\upmu$s, the oscillation frequency jumps from $\sim 1.7\,$MHz to $\sim 3.5\,$MHz corresponding to a state change from G- to C-state. With the local oscillator at $339.3\,$MHz,  the related real frequencies at $I_{dc}=10.25\,$mA can consequently be identified as $341\,$MHz for the predominant G-state and $335.8\,$MHz for the C-state. Note that these frequency differences are not properly resolved in the frequency determination by a spectrum analyzer as shown in fig. \ref{fig_C-state:spectra_RTN} (typical measurement duration $\sim 10\,$ms).  
The probability to be in the C-state can now be analyzed and is  shown in fig. \ref{fig_C-state:probability} through the transition range. Its evolution describes a classical stochastic phase transition with the dc current as control parameter. 
We recognize that a $50\,$\% probability is reached at $\sim 10.6\,$mA, the current value where also the slope of the characteristic RTN frequency evolution is roughly estimated to invert (see fig. \ref{fig:RTN-fit-parameters}, disregarding points with large errors between $10$ and $10.6\,$mA). 
In general, random telegraph noise is a major limitation especially for magnetic sensor applications \cite{Almeida2008,Weitensfelder2018} and can be removed here by stabilization of the G- or C-mode.  
However, in the sense of the G- to C-state transition, it can serve as a sort of easily accesible indicator of the probabilistic properties.

 \section{Discussion \& Conclusion}

 We present here measurements of the dynamic C-state in vortex based spin torque nano oscillators and analyze the system's phase noise characteristics. 
 Regarding the evolution of the system's flicker noise, we find its increase with increasing applied dc current in the measured C-state regime that is dominated by Hooge's law. 
 The measurement confirms that a minimum in the flicker noise level exists also in the G-state when the active oscillation volume  saturates with the current. 
  
 At the transition between G- and C-state, we demonstrate the system oscillating for several periods in the G- and several periods in the C-state, stochastically switching between both states at a characteristic frequency in the kHz range. 
 The occurrence of random telegraph noise in the stochastic regime is demonstrated. 
 Its determined characteristic parameters are compared with the state probability evaluated from a time-frequency analysis and can serve as a sort of  indicator of the stochastics. The parameters show a controllable dependence of the probabilistic properties with the applied current. 
 
 We believe, firstly, that the presented results bring more insight  into the noise characteristics of spin torque nano oscillators, specifically those based on the texture of a magnetic vortex, but also partly generalizable to other types of STOs. 
 Secondly, the characterization of the stochastic transition between the two dynamic states in vortex based spin torque oscillators is an important step in order to add another feature to the exploitation of STVOs, namely stochasticity, demonstrating the STVO's capability of real multifunctionality sought in multiple technologies, such as neuromorphic systems.


\begin{acknowledgments}
Gilles Cibiel, Serge Galliou and Enrico Rubiola are acknowledged for fruitful discussions. 
S.W. acknowledges funding from Labex FIRST-TF under contract number ANR-10-LABX-48-01. The work is supported by the French ANR project ''SPINNET'' ANR-18-CE24-0012. 
P.T. acknowledges   support under   the   Cooperative   Research   Agreement   Award No.  70NANB14H209,  through  the  University  of  Maryland. 
M.R. acknowledges support from Spanish MICINN (PGC2018-099422-A-I00) and Comunidad de Madrid (Atracción de Talento Grant No. 2018-T1/IND-11935). 

\end{acknowledgments}

\section*{\footnotesize{Data Availability Statement}}

The data that support the findings of this study are available from the corresponding author upon reasonable request.

\appendix

\bibliography{literatur_promo}

\end{document}